\documentclass[aps,prl,twocolumn,groupedaddress]{revtex4-1}
\usepackage{graphicx}
\usepackage{dcolumn}

\begin{document}
\newcommand{\mnras}{Mon. Not. R. Astron. Soc.}
\newcommand{\apjl}{Astrophys. J. Lett.}
\newcommand{\grl}{Geophys. Res. Lett.}
\newcommand{\jgr}{J. Geophys. Res.}
\newcommand{\apjs}{Astrophys. J. Supp.}
\newcommand{\aap}{Astron. Astrophys.}
\newcommand{\apss}{Astrophys. Space Sci.}


\title{Magnetic Helicity Conservation and Inverse Energy Cascade \\
     in Electron Magnetohydrodynamic Wave Packets}


\author{Jungyeon Cho}
\email[]{jcho@cnu.ac.kr}
\altaffiliation{Dept. of Astronomy \& Space Science, Chungnam National Univ., Daejeon, Korea}


\date{\today}

\begin{abstract}
Electron magnetohydrodynamics (EMHD) provides a fluid-like description of
small-scale magnetized plasmas.
An EMHD wave (also known as whistler wave) propagates along magnetic field lines.
The direction of propagation can be either parallel or anti-parallel to
the magnetic field lines.
We numerically study propagation of 3-dimensional (3D) EMHD
wave packets moving in one direction. 
We obtain two major results: 1. Unlike its magnetohydrodynamic (MHD)
counterpart, an EMHD wave packet is dispersive. Because of this, EMHD
wave packets traveling in one direction create opposite traveling wave packets
via self-interaction
and cascade energy to smaller scales.
2. EMHD wave packets traveling in one direction clearly exhibit inverse energy cascade.
We find that the latter is due to conservation of magnetic helicity.
We compare inverse energy cascade in 3D EMHD turbulence
and 2-dimensional (2D) hydrodynamic turbulence.
\end{abstract}




\maketitle

\section{1. Simulations and Results}

{\bf Introduction ---} Astrophysical plasmas are observed in a wide range of length-scales.
On large scales, we can treat such plasmas as conducting fluids and therefore
we can use magnetohydrodynamics (MHD).
Although MHD is a simple and powerful tool for large scales, it is not suitable for
describing small-scale physics, especially physics near and below the proton
gyro-scale.
Since many astrophysical processes critically depend on small-scale physics,
proper description of small-scale physics is needed.
There are several numerical models that can handle small-scale physics.
Perhaps a full kinetic treatment would be the best for proper description of
small-scale physics. However, 
a full kinetic description of plasmas is still a challenge for 
modern computers.
 
Electron magnetohydrodynamics (EMHD) is a fluid-like model of
small-scale plasmas\cite{KinCY90}
and can be viewed as
Hall MHD in the limit of $k\rho_i \gg 1$, where
$\rho_i$ is the ion gyroradius and $k$ the wavenumber.
On scales below the ion inertial length $d_{i}=c/\omega_{pi}$, where
$c$ is the speed of light and $\omega_{pi}$ is the ion plasma frequency,
we can assume that
the ions create only smooth motionless background and fast electron flows
carry all the current, so that
\begin{equation}
  {\bf v}_e 
   = - \frac{ {\bf J} }{ n_ee } 
            = - \frac{ c }{ 4 \pi n_e e } \nabla \times {\bf B}, \label{eq1}
\end{equation}
where ${\bf v}_e$ is the electron velocity, {\textbf J} is the electric current density,
{\textbf B} is the magnetic field, $n_e$ is the electron
number density, and $e$ is the absolute value of the electric charge.
Inserting this into the usual magnetic induction equation
($\partial {\bf B}/\partial t = \nabla \times ({\bf v}_e \times {\bf B})+ \eta \nabla^{2} {\bf B}$),
we obtain the EMHD equation
\begin{equation}
   \frac{ \partial {\bf B} }{ \partial t }
  = - \frac{ c }{ 4 \pi n_e e } \nabla \times \left[
    (\nabla \times {\bf B}) \times{\bf B} \right] + \eta \nabla^{2} {\bf B}.
\end{equation}
Note that, in this paper, we only consider the zero (normalized)
electron inertial length case: $d_e=c/(\omega_{pe}L) \rightarrow 0$,
where $\omega_{pe}$ is the electron plasma frequency and $L$ is the
typical size of the system.

As is the case with large scales, magnetized turbulence on small scales also
affects many physical processes and hence is of great interest for studies of magnetic reconnection\cite{DraKM94, Bisk96},
space plasmas and
the solar wind\cite{DmiM06pp, GarSL08,  HowDC08, HowCD08,  LeaSN98, MatSD08, SaiGL08, SchCD09, SmiHV06, StaGL01},
neutron stars\cite{GolR92,CumAZ04,WarH09aa},
advection dominated 
accretion flows\cite{QuaG99}, 
etc.
Due to its simplicity, EMHD formalism has been used for studies 
of small-scale turbulence\cite{BisSD96,BisSZ99,NgBG03,ChoL04,ShaZ05,ChoL09,WarH09pp,WarH10}.
Earlier studies revealed that
$E_b(k) \propto k^{-7/3}$\cite{Vai73,BisSD96,BisSZ99}, and anisotropic turbulence structures\cite{DasDK00,ChoL04,ChoL09}.

In the presence of a strong mean field ${\bf B}_0$,
an MHD perturbation moves along magnetic field at the Alf\'{v}en speed ($\propto B_0$) and
MHD wave packets moving
in one direction do not interact each other and do not create turbulence.
Therefore, collisions of opposite-traveling wave packets are essential for generation of MHD turbulence.
In contrast, an EMHD perturbation moves along magnetic field at a speed proportional to $kB_0$,
which
implies that a perturbation with larger $k$ is faster than that with smaller $k$.
As a result, whistler waves are dispersive and whistler wave packets moving in one direction
can self-interact and produce small-scale structures (see \cite{NgBG03} for 2D EMHD),
which means that collisions of whistler wave packets are not essential for generation of EMHD turbulence.
Therefore understanding the dynamics of EMHD wave packets is important for study of EMHD turbulence.
Traveling EMHD wave packets can commonly occur in nature. 
Any local disturbances can create wave packets traveling along magnetic field lines.
Therefore, propagation of whistler wave packets moving in one direction deserves a scrutiny.
In this paper, we study propagation of 3D EMHD wave packets in detail.

{\bf Numerical Method ---} We have calculated the time evolution of 
3D incompressible EMHD wave packets moving in one direction.
We have adopted a pseudospectral code
to solve the normalized incompressible EMHD equation in a
periodic box of size $2\pi$:
\begin{equation}
\frac{\partial {\bf B}}{\partial t}=-
     \nabla \times \left[
    (\nabla \times {\bf B}) \times{\bf B} \right] 
    + \eta^{\prime} \nabla^{2} {\bf B},
     \label{beq}
\end{equation}
where the magnetic field, time, and length are normalized by
a mean field $B_0$, the whistler time $t_w=L^2(\omega_{pe}/c)^2/\Omega_e$
($\Omega_e$= electron gyrofrequency), and
a characteristic length scale $L$ 
(see, e.g., \cite{GalB03}).   
The resistivity $\eta^{\prime}$ in equation (\ref{beq}) is dimensionless.
The dispersion relation of a whistler wave in this normalized unit
is $\omega=kk_{\|}B_0$, where $k_{\|}$ is the wave number parallel
to the 
mean magnetic field.
The magnetic field consists of the uniform background field and a
fluctuating field: ${\bf B}= {\bf B}_0 + {\bf b}$.
The strength of
the uniform background field, $B_0$, is set to 1.
We use either $256^3$ or $512^3$ collocation points.
At $t=0$, all waves are moving in the same direction and their wave numbers
are restricted to the range
$8\leq k \leq 15$ in wavevector ({\bf k}) space.
The direction of propagation corresponds to
the {\it positive} direction of the mean magnetic field in our simulations.
(Hereinafter, we use {\it positive} and {\it negative} to denote
the direction of wave propagation with respect to the magnetic field.)
The amplitudes of the random magnetic field at $t=0$ is $\sim 1$.
Hyperdiffusivity is
used for the diffusion term.
The power of hyperdiffusivity
is set to 3, so that the dissipation term in the above equation
is replaced with
  $\eta_3 (\nabla^2)^3 {\bf B}$,
where $\eta_3$ is approximately $3 \times 10^{-10}$ for $256^3$ and 
$1\times 10^{-11}$ for $512^3$. 

\begin{figure*}
\includegraphics[angle=0,width=0.42\textwidth]{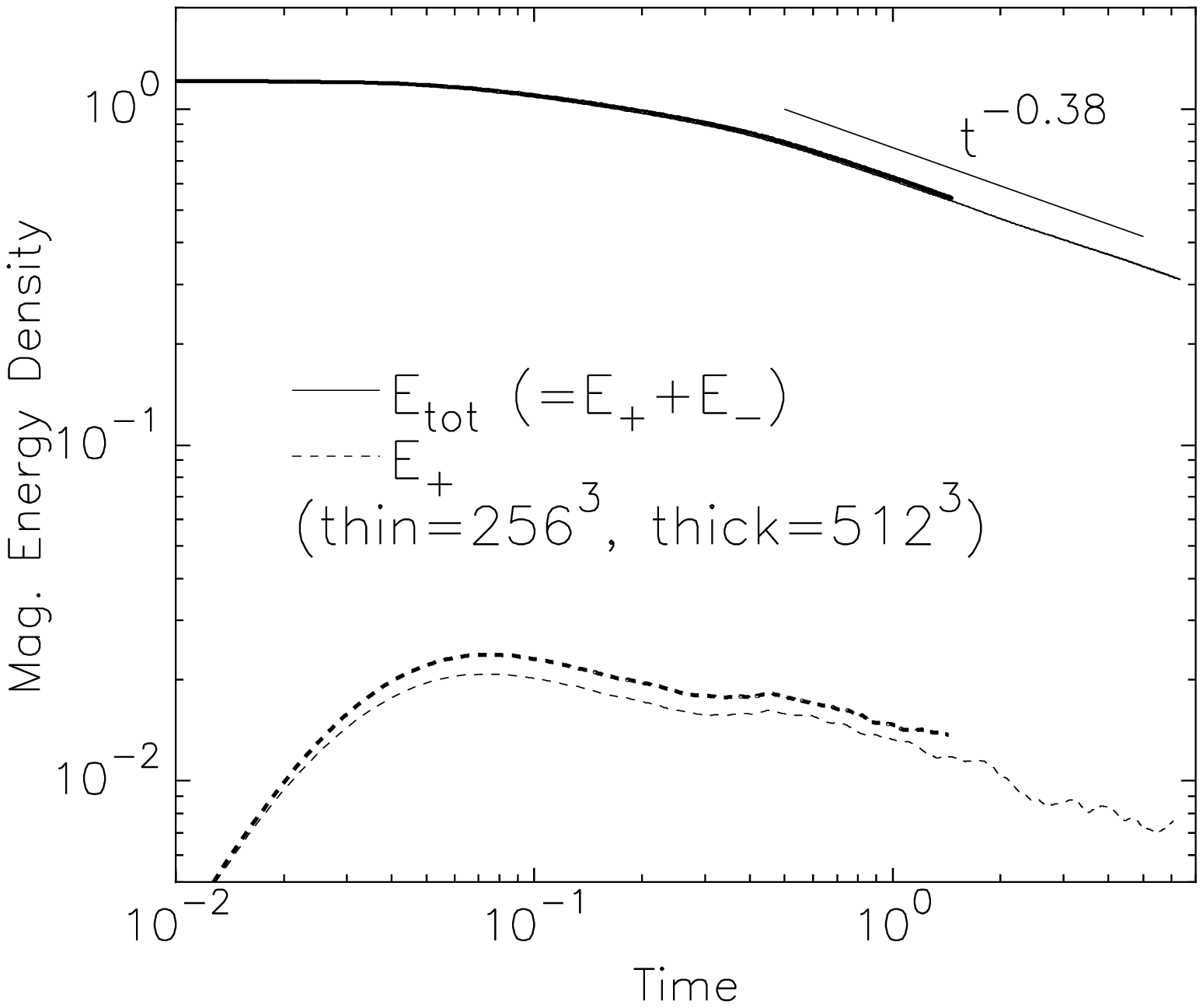}
\hspace{10mm}
\includegraphics[angle=0,width=0.42\textwidth]{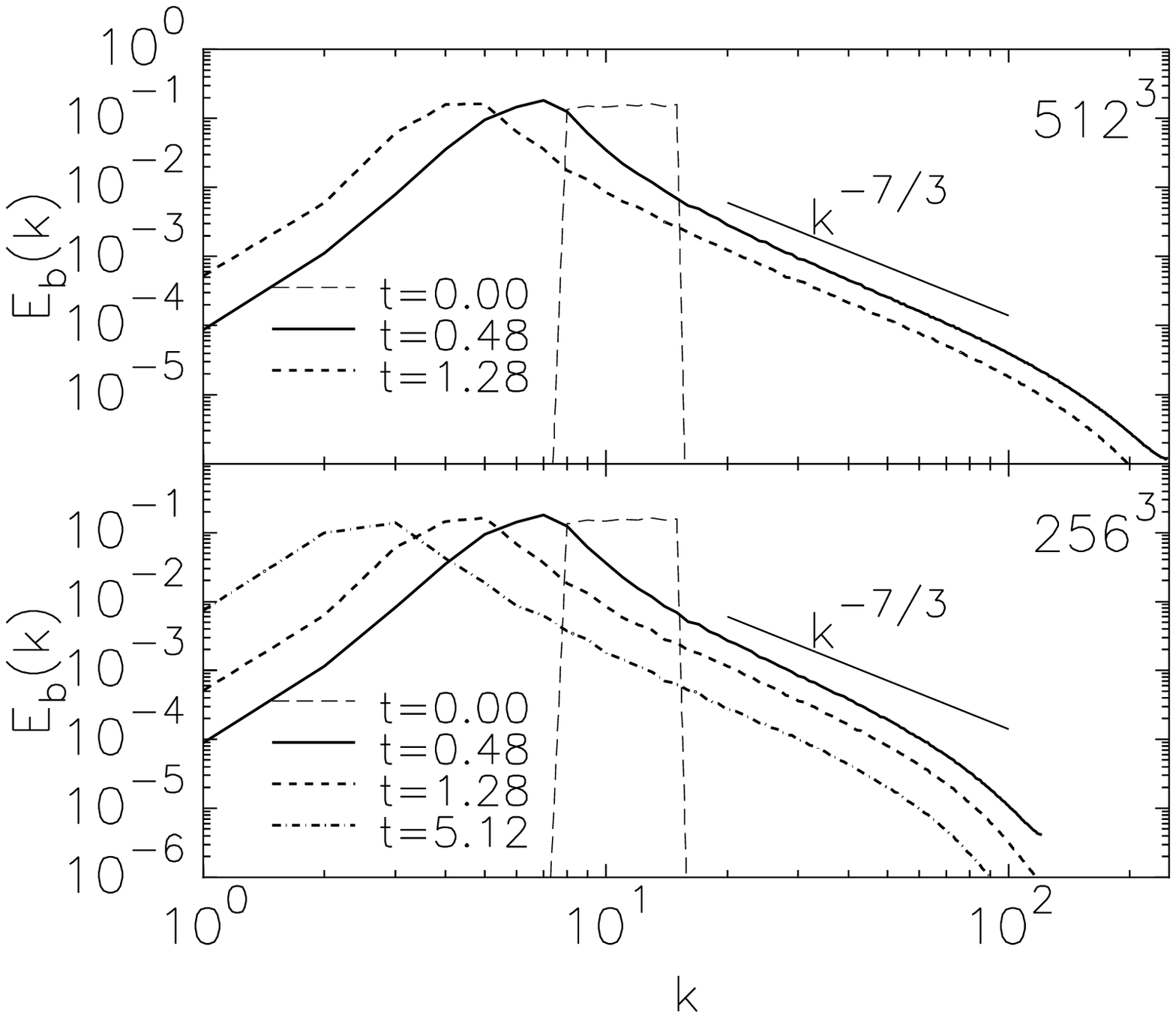}
\caption{ Time evolution of the fluctuating magnetic energy density $b^2$. 
          Initially all wave packets move in the same direction.
          Due to self-interaction, $b^2$ decreases as time goes on (solid curves).
          Self-interaction generates waves traveling in the opposite direction (dotted curves).
          The energy of the opposite-traveling waves is no larger than a few \% of
          the energy of the waves moving in the original direction.
          The thin curves are for $256^3$ and the thick ones for $512^3$.
          \label{fig1} }
\caption{ Magnetic spectra showing inverse energy cascade.
         The peak of magnetic energy spectrum moves to larger scales.
         Note that $E(k_p)$, where $k_p$ is the wave number at which the magnetic energy spectrum peaks,
         is almost constant.
         Magnetic helicity conservation plays a key role in the inverse cascade.   \label{fig2}}
\end{figure*}


{\bf Results ---} Figure~\ref{fig1} shows time evolution of magnetic
energy density. 
The solid curves denote the total magnetic energy density (i.e. the total energy
of the wave packets moving in both the positive and the negative directions) and
the dotted curves the energy of wave packets moving in the negative direction.
We can see that wave packets initially moving only in the positive direction 
can create waves moving in the negative direction.
Although the energy of wave packets moving in the negative direction is small 
(at most a few \% compared with the energy of the wave packets moving in the positive direction),
we clearly observe that magnetic energy decays.
Results from $512^3$ (thick curves) and $256^3$ (thin cures) show a reasonable agreement.

Figure~\ref{fig2} shows magnetic energy spectrum as a function of time. 
At $t=0$, Fourier modes between $k=8$ and $k=15$ 
are excited. The long-dashed curves in Figure~\ref{fig2} show the
initial spectrum.
As time goes on, the initial energy cascades down to smaller scales and, as a result, 
a power-law-like spectrum forms for $k>15$.
At the same time the peak of the energy spectrum moves to larger scales, so that
the wavenumber at which the spectrum peaks, $k_p$, gets smaller.
We clearly observe inverse cascade of magnetic energy.

\begin{figure}
\includegraphics[angle=0,width=0.42\textwidth]{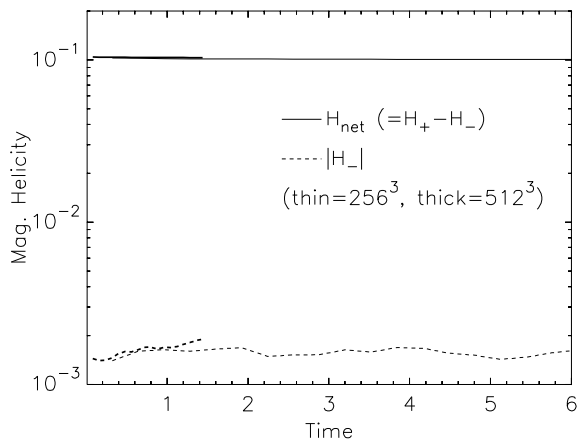}
\caption{Helicity conservation.
         The net magnetic helicity (solid curves) is well-conserved.
         The dotted curves denote the absolute values of magnetic helicity in opposite-traveling wave packets,
         which are generated by self-interaction of the initial wave packets. 
         The thin curves are for $256^3$ and the thick ones for $512^3$. \label{fig3}}
\end{figure}

The magnetic helicity is a conserved quantity in EMHD\cite{GalB03}.  
Figure~\ref{fig3} shows that the magnetic helicity is extremely well conserved.
The solid curves denote the net magnetic helicity (i.e. the helicity of wave packets moving
in the positive direction minus that in the negative direction)
and 
the dotted curves the helicity of wave packets moving in the negative direction.
The helicity of the wave packets moving in the negative direction is much
smaller than that of the waves moving the other way. 
As in Figure~\ref{fig1}, thick curves are for $512^3$ and thin curves for $256^3$.

\begin{figure}
\includegraphics[angle=0,width=0.22\textwidth]{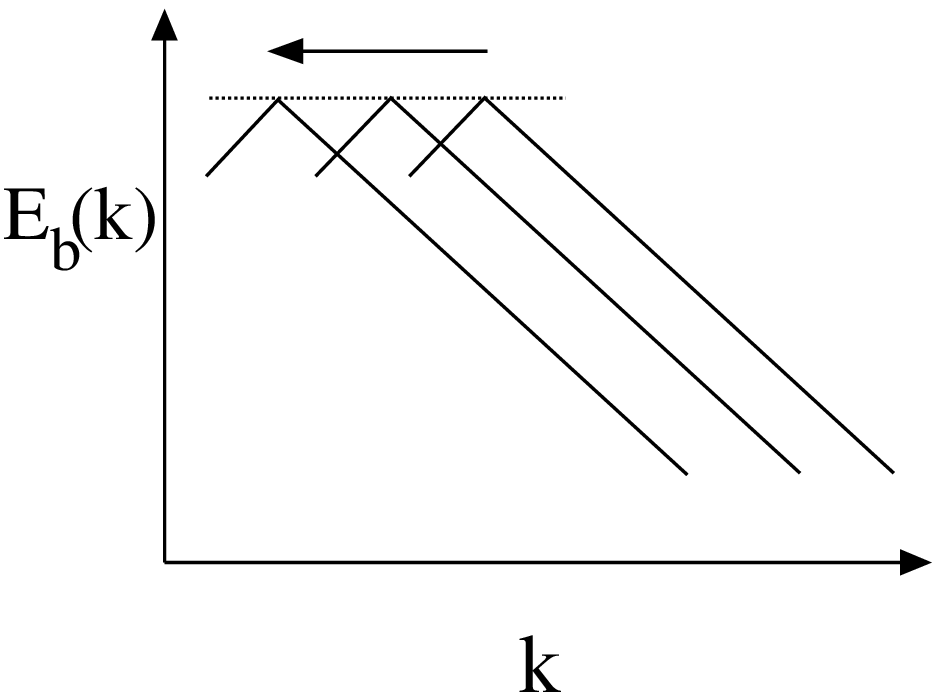}  
\includegraphics[angle=0,width=0.23\textwidth]{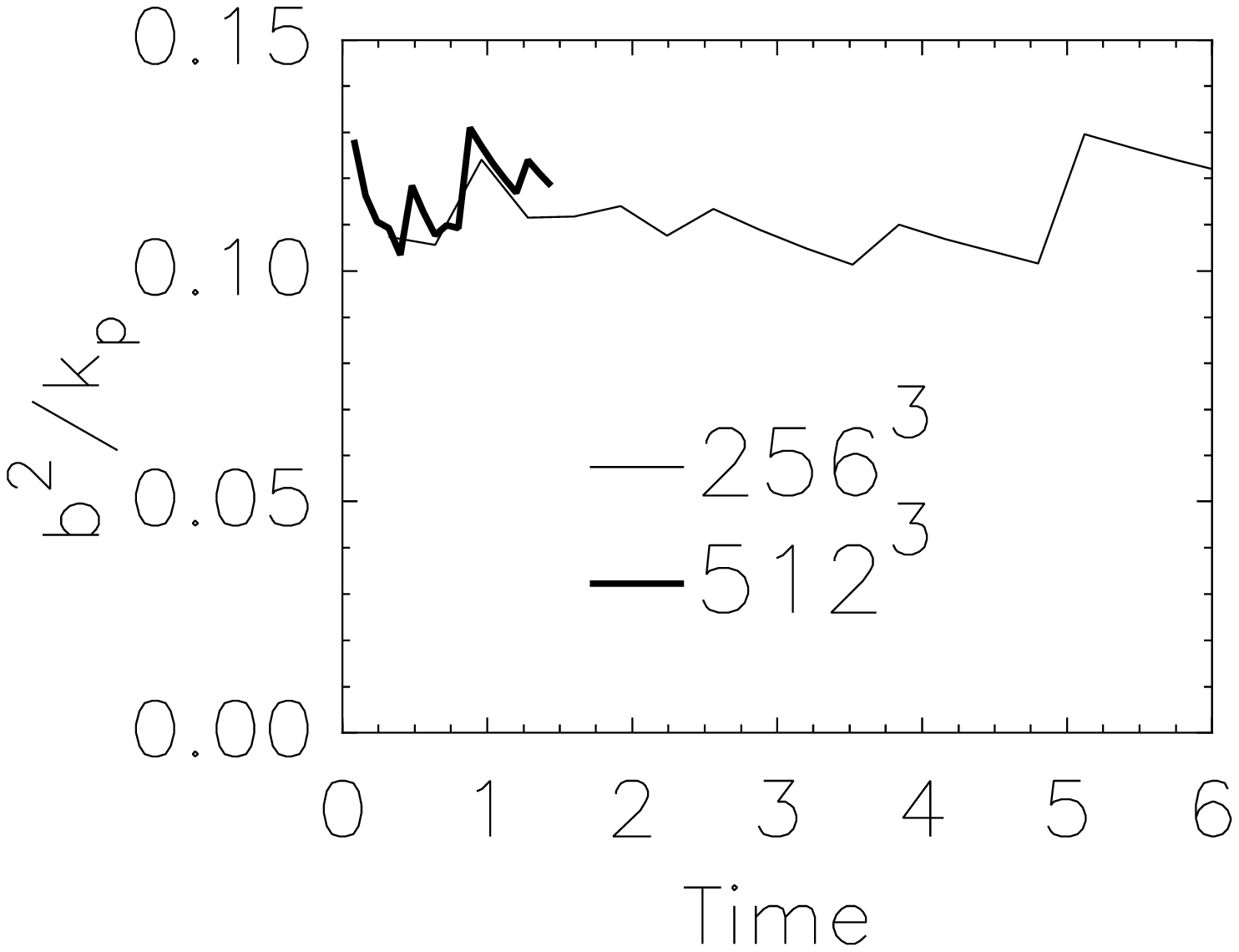}  
\caption{ Quantities that remain constant during inverse energy cascade.
          We expect that $E_b(k_p)$ and $b^2(t)/k_p(t)$ are almost constant.
          {\it Left panel:} Schematic diagram showing 
            constancy of $E_b(k_p)$.            
          See Figure~\ref{fig2} for actual simulation results.  
          {\it Right panel:} Simulation result that confirms constancy of $b^2(t)/k_p(t)$.
         %
                 \label{fig4}}
\end{figure}

\section{2. Theoretical Interpretations}

{\bf Helicity decomposition ---} 
An EMHD wave traveling along the magnetic field lines is circularly polarized.
In Fourier space, the bases that describe circular polarization are
\begin{equation}
    \hat{\bf \epsilon}_{+} \equiv (\hat{\bf s}_1 + i \hat{\bf s}_2 )/\sqrt{2} \mbox{ ~~and~} 
    \hat{\bf \epsilon}_{-} \equiv (\hat{\bf s}_1 - i \hat{\bf s}_2 )/\sqrt{2},
\end{equation}
where $\hat{\bf s}_1$ and $\hat{\bf s}_2$ are orthogonal unit vectors perpendicular to ${\bf k}$ and
we assume $\hat{\bf s}_1 \times \hat{\bf s}_2 = \hat{\bf k} = {\bf k}/k$.
Any EMHD Fourier mode can be decomposed into a `+' wave and a `-' wave:
\begin{equation}
  \tilde{\bf b}_{\bf k} = \tilde{b}_{+} \hat{\bf \epsilon}_{+}
                + \tilde{b}_{-} \hat{\bf \epsilon}_{-},
\end{equation}
where $\tilde{\bf b}_{\bf k}=\tilde{b}_1 \hat{\bf s}_1 +\tilde{b}_2 \hat{\bf s}_2$, 
   $\tilde{b}_{+}=(\tilde{b}_1 -i \tilde{b}_2)/\sqrt{2}$, and 
   $\tilde{b}_{-}=(\tilde{b}_1 +i \tilde{b}_2)/\sqrt{2}$.

The `+' wave moves in the {\it positive} direction and 
the `-' wave in the {\it negative} (i.e.~apposite) direction with respect to magnetic field.
Let us consider a `+' wave.
In Fourier space, we can show that
\begin{equation}
   \partial (\tilde{b}_+ \hat{\bf \epsilon}_+) / \partial t = -i kk_{\|}B_0 (\tilde{b}_+ \hat{\bf \epsilon}_+ )
\end{equation}
(see, e.g.~\cite{GalB03}).
Therefore, we have $\tilde{b}_+ = b_0\exp(-i\omega t)$, where $b_0$ is a constant.
Fourier transform of $\tilde{b}_+$ to real space gives
a plane wave whose velocity relative to the magnetic field is positive. 

The `+' wave has a positive magnetic helicity and
the `-' wave a negative magnetic helicity.
Let us consider a `+' wave.
In Fourier space, we can show that the Fourier component of the vector potential is given by
\begin{equation}
   \tilde{\bf a}_{\bf k}=\tilde{a}_+ \hat{\bf \epsilon}_+ = ( \tilde{b}_+ /k )\hat{\bf \epsilon}_+,
\end{equation}
where we use the Coulomb gauge.
Therefore, the magnetic helicity of the `+' wave is positive.

{\bf Inverse cascade ---}
The magnetic helicity is a conserved quantity in EMHD (see, e.g. \cite{GalB03}).  
For an EMHD wave packet composed of only `+' waves,
the magnetic helicity spectrum $E_h(k)$ is simply
\begin{equation}
   E_h(k) = E_b(k)/k.  \label{eq_heli}
\end{equation}
Therefore, the helicity dissipation rate becomes negligible
when $k_d$, where $k_d$ is the dissipation wavenumber, is large with respect to $k_p$.
Note, however, that energy dissipation rate can be non-negligible.
Therefore, as energy dissipates, the peak wavenumber $k_p$ should become smaller.

Since $\int E_h(k) dk$ is nearly constant 
and $\int E_h(k) dk \sim [E_b(k_p) /k_p ] k_p \sim E_b(k_p)$, 
we have $E_b(k_p)\approx$constant
(see Figure 2; see also left panel of Figure~\ref{fig4}).   
Since $b^2 \sim E_b(k_p)k_p \propto k_p$, we expect that
\begin{equation}
    k_p(t) \propto b^2(t), 
\end{equation}
which is confirmed by   
right panel of Figure~\ref{fig4}.
We obtained $k_p$  in such a way that 
$\sum_{ k-0.5 \leq |{\bf k}^\prime| < k+0.5} | \tilde{\bf b}({\bf k}^\prime )|^2$ 
($k=1.0, 1.2, 1.4, 1.6, ...$)
is maximum at $k=k_p$.

{\bf Comparison with 2D hydrodynamic turbulence ---}
We find a very good correspondence between 3D EMHD turbulence
and 2D incompressible hydrodynamic turbulence.
The 2D incompressible hydrodynamic equation has two ideal invariants: 
the energy and the enstropy ($\propto \int |\nabla \times {\bf v}|^2 d^2 {\bf x}$).
The spectrum of enstropy is 
   $ E_{en}(k) = k^2 E_v(k)$,  
where $E_v(k)$ is energy spectrum of velocity.
In driven 2D hydrodynamic turbulence, both energy and enstropy involve turbulence cascade:
enstropy exhibits forward cascade and energy
inverse cascade.
When energy is the cascading quantity, which is the case for scales larger than the energy injection
scale, we have 
\begin{equation}
   \frac{ v_l^2 }{ l/v_l } \sim \frac{ kE_v(k) }{ 1/(k \sqrt{kE_v(k)} } \mbox{=const.} 
   \rightarrow E_v(k) \propto k^{-5/3},
\end{equation}
where we used $v_l \sim \sqrt{kE_v(k)}$.
On the other hand, when enstropy is the cascading quantity, 
which is the case for scales smaller than the energy injection scale, we have
\begin{equation}
    \frac{ kk^2E_v(k) }{ 1/(k \sqrt{kE_v(k)} } \mbox{=const.} 
   \rightarrow E_v(k) \propto k^{-3},
\end{equation}
where we used $E_{en}(k) = k^2 E_v(k)$.   
In this case, interactions are non-local in wavevector space.

In 3D EMHD turbulence, magnetic energy and magnetic helicity are conserved quantities
in the absence of dissipation.
Therefore, either magnetic energy or magnetic helicity can involve energy cascade.
Using arguments similar to the 2D hydrodynamic case, we can obtain
magnetic energy spectrum.
When magnetic energy is the cascading quantity, we have
\begin{equation}
   \frac{ b_l^2 }{ l/v_l } \propto \frac{ kE_b(k) }{ 1/(k \sqrt{kk^2E_b(k)} } \mbox{=const.} 
   \rightarrow E_b(k) \propto k^{-7/3},
\end{equation}
where we used ${\bf v} \propto {\bf J} \propto \nabla \times {\bf B}$, which means
$E_v(k)\propto k^2 E_b(k)$.
On the other hand, when magnetic helicity is the cascading quantity, 
we have
\begin{equation}
    \frac{ kE_b(k)/k }{ 1/(k \sqrt{kk^2E_b(k)} } \mbox{=const.} 
   \rightarrow E_b(k) \propto k^{-5/3},
\end{equation}
where we used Eq.~(\ref{eq_heli}).

When we inject magnetic helicity (and, hence, magnetic energy) on a scale, we expect to see 
both inverse and forward
energy cascade.
It is well-known that when energy is injected on a scale, the energy
cascades down to smaller scales and the small-scale magnetic energy spectrum 
is proportional to $k^{-7/3}$.
Therefore, it is evident that magnetic energy exhibits forward cascade.
However, it is not clear what is the cascading entity for inverse cascade. 
We will address this issue elsewhere (Kim et al., in preparation).

In summary, we have shown that EMHD wave packets moving in one direction can create opposite-traveling
      wave packets through self-interaction and that,
      because of magnetic helicity conservation, EMHD wave packets moving in one direction
      show inverse energy cascade.
Inverse cascade of energy can affect transport phenomena, such as heat transport, and
potentially be a source of plasma instabilities.
It is also potentially important for magnetic reconnection and evolution of magnetic field in neutron stars.
In general, traveling EMHD wave packets can be a source of magnetic helicity, which could affect
evolution of a large-scale magnetic field.

\begin{acknowledgments}
This research was supported by National R\&D Program through 
the National Research Foundation of Korea(NRF) 
funded by the Ministry of Education, Science and Technology (No. 2010-0029102).
JC thanks H. Kim for useful discussions.
\end{acknowledgments}

\bibliographystyle{apsrev4-1}
\bibliography{ref}

\end{document}